\documentclass[aip,apl,reprint,superscriptaddress]{revtex4-2}
\usepackage{graphicx}
\usepackage{amsmath}
\usepackage{amssymb}
\usepackage{natbib}

\begin{document}

\title{Perspective on 2D perovskite ferroelectrics and multiferroics}

\author{Junting Zhang}
\email{juntingzhang@cumt.edu.cn}
\thanks{Authors to whom correspondence should be addressed}
\author{Yu Xie}
\author{Ke Ji}
\affiliation{School of Materials Science and Physics, China University of Mining and Technology, Xuzhou 221116, China}
\author{Xiaofan Shen}
\email{xiaofanshen@smail.nju.edu.cn}
\affiliation{National Laboratory of Solid State Microstructures and Physics School, Nanjing University, Nanjing 210093, China}

\begin{abstract}
Two-dimensional (2D) ferroelectrics and multiferroics have attracted considerable scientific and technological interest in recent years due to the increasing demands for miniaturization and low energy consumption of electronic devices. At present, the research on 2D ferroelectrics and multiferroics is still focused on van der Waals materials, while the known bulk ferroelectric and multiferroic materials are mostly found in perovskite systems. The ability to prepare and transfer 2D perovskite oxides has provided unprecedented opportunities for developing ferroelectrics and multiferroics based on 2D perovskites. In this Perspective, we review the research progress on 2D ferroelectrics and multiferroics in inorganic perovskites in terms of different ferroelectric and magnetoelectric coupling mechanisms. The improper ferroelectricity and novel magnetoelectric coupling mechanisms discovered in 2D perovskites are emphasized, and then the main challenges and future development direction are put forward.
\end{abstract}
\maketitle

\section{Introduction}

Research on ferroelectricity has flourished for a century since it was first discovered in Rochelle salt. \cite{Ramesh2001,Scott2007,Nuraje2013} At present, many types of ferroelectrics have been discovered, among which perovskite ferroelectrics are the most extensively studied system. \cite{Goodenough2004,Bokov2006,Zheng2023,Fu2021} The perovskite family with chemical formula $ABX_3$, where \emph{A} and \emph{B} are cations and \emph{X} is an anion, has thousands of candidate materials due to the rich combination of cations. However, conventional perovskite ferroelectrics usually require cations to have a special electronic configuration, that is, either the \emph{A}-site cation has stereochemically active lone pair electrons (such as Bi$^{3+}$ and Pb$^{2+}$ ions) or the \emph{B}-site transition-metal ion has a $d^0$ configuration (such as Ti$^{4+}$ and Nb$^{5+}$ ions), which results in a small amount of perovskite ferroelectrics. \cite{Bersuker2013,Benedek2013,Qiao2021,Dawber2005} This proper ferroelectricity has been extensively identified and attributed to the second-order Jahn-Teller (JT) effect. \cite{Bersuker2013}

Reducing the thickness of ferroelectric films to nanometer size has always been a key challenge in the research of ferroelectricity. \cite{Qiao2021,Dawber2005,Shaw2000,Ahn2004} Conventional perovskite ferroelectrics commonly suffer from the critical thickness effect, that is, their ferroelectricity disappears when reduced to tens or a few nanometers in thickness. \cite{Petkov2008,Junquera2003,Fong2004,Kim2005,Woo2007} This is due to the increase of the depolarization field caused by the reduction of thickness, \cite{Woo2007} which leads to the instability of the ferroelectric phase. However, there have been some indications that the critical thickness may be absent in ultrathin perovskite ferroelectric films. \cite{Kolpak2005,Gao2017,Jia2006} Therefore, whether ferroelectricity can be retained to the 2D limit, has always been a controversial issue. \cite{Qiao2021} Actually, 2D ferroelectricity was first demonstrated in van der Waals (vdW) materials, where some ferroelectric monolayers exhibit out-of-plane (OP) ferroelectricity. \cite{Chang2016,Liu2016,Ding2017,Zheng2018,Zhou2017,Qi2021,Wu2018} In addition, some novel ferroelectric phenomena have been discovered in the research on 2D vdW ferroelectrics, such as the emergence of ferroelectricity in elemental monolayers and stacked bilayers, and fractional quantum ferroelectricity. \cite{Li2017,Xiao2018,Fei2018,Gou2023,Li2023,Wang2023b,Ji2023,Ji2024}

In addition to ferroelectricity, perovskite systems also exhibit rich correlated electronic behaviors, such as superconductivity, giant magnetoresistance, and multiferroicity. \cite{Dagotto2005,Dagotto2001,Eerenstein2006} Multiferroic is defined as having two or more ferroic orders in a single material, involving ferroelectricity, ferromagnetism, and ferroelasticity. Multiferroic materials may exhibit new functions due to the coupling of different ferroic order parameters, such as magnetoelectric coupling effect. \cite{Eerenstein2006,Ramesh2007,Cheong2007,Tokura2010,Tokura2014,Dong2015} The cross control of ferroic order parameters by external fields can be achieved based on the magnetoelectric coupling, which is of great significance for the development of next-generation electronic devices with high density, high processing speed, multifunctionality, and low energy consumption. \cite{Catalan2009,Bibes2008,Chu2008,Heron2014,Spaldin2019} The resurgence of research on multiferroics began with perovskites, which can be traced back to the discovery of two representative multiferroic materials, BiFeO$_3$ and TbMnO$_3$, in 2003. \cite{Wang2003,Kimura2003,Fiebig2005,Spaldin2005,Fiebig2016} The former has proper ferroelectricity and is classified as the type-I multiferroics, while the latter has spin-induced ferroelectricity and is called the type-II multiferroics.

Since 2D ferroelectricity and ferromagnetism were demonstrated in vdW materials, 2D multiferroics have attracted increasing interests. \cite{Luo2017,Huang2018,Qi2018,Lai2019,Tan2019,Ai2019,You2020,Liu2020,Zhang2018,Song2022,Chu2020} At present, the research on 2D multiferroics mainly focuses on vdW materials. Although 2D multiferroics with coexisting ferroelectricity and ferromagnetism have been predicted and achieved in some vdW monolayers, they rarely exhibit intrinsic magnetoelectric coupling effects. \cite{Gao2021,Tang2019,Behera2021,Du2022,Tang2023,Xun2024}

\begin{figure*}[htbp]
\centering
\includegraphics*[width=0.85\textwidth]{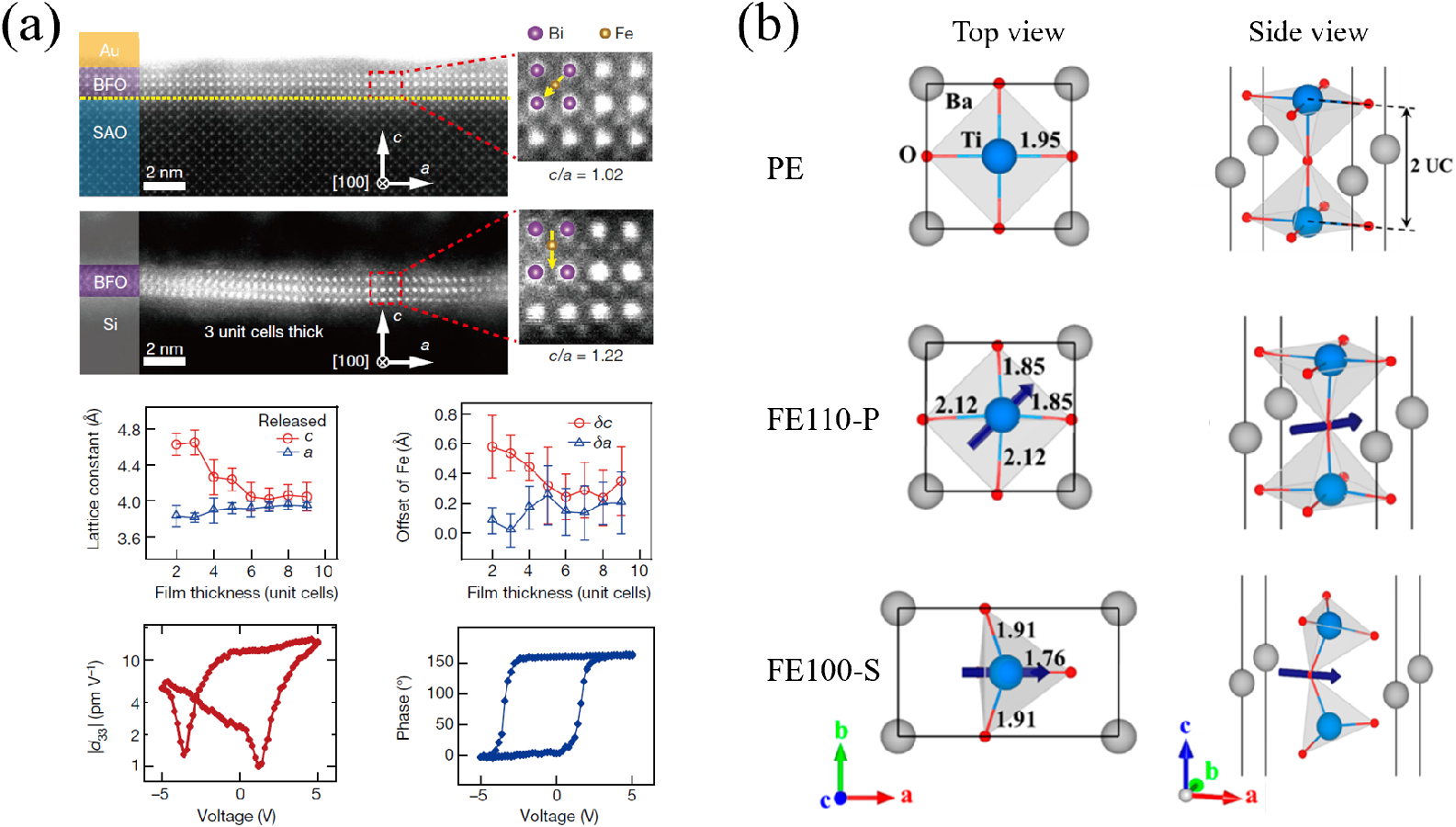}
\caption{\label{Fig1} (a) OP ferroelectricity and lattice distortion in freestanding 2D BiFeO$_3$ films. Reproduced with permission from Ji \emph{et al.}, Nature, 570, 87 (2019). Copyright 2019 Springer Nature. \cite{Ji2019} (b) Ferroelectricity in 2D BaTiO$_3$ films with TiO$_2$ termination. The upper, middle, and lower panels show the top and side views of the geometrical structures of the paraelectric phase, proper ferroelectric phase, and ferroelectric phase induced by surface effects, respectively. Reproduced with permission from Lu \emph{et al.}, Nano Lett., 18, 595 (2018). Copyright 2019 American Chemical Society. \cite{Lu2017}}
\end{figure*}

In recent years, some breakthroughs have been made in the research on 2D ferroelectricity and multiferroics in perovskites. Experimentally, the freestanding perovskite oxide films down to the monolayer limit have been successfully prepared and can be transferred to any desired substrate, opening the door to exploring functional properties based on 2D perovskites. \cite{Lu2016,Hong2017,Ji2019,Ricciardulli2021} Theoretical studies show that the proper ferroelectricity that occurs in conventional ferroelectrics can be retained at the 2D limit. \cite{Lu2017,Zhang2024} Some improper ferroelectric mechanisms have also been found in 2D perovskites, involving octahedral distortion induced ferroelectricity and spin-induced ferroelectricity, based on which the coexistence and coupling of ferroelectricity and ferromagnetism can be achieved. \cite{Zhou2022,Zhou2021,Shen2023,Zhang2020,Shen2021,Zhang2022} Importantly, some novel magnetoelectric coupling effects that are lacking in perovskite bulks can emerge in 2D perovskites, promising the magnetoelectric cross control involving the control of magnetization by electric field and polarization by magnetic field. \cite{Zhou2021,Shen2023,Zhang2020,Shen2021,Zhang2022} Therefore, 2D perovskites provide a broad platform for designing and exploring 2D ferroelectricity and magnetoelectric multiferroics.

In this Perspective, the background and research progress of 2D ferroelectricity and multiferroicity in inorganic perovskites are briefly reviewed. The remainder of this Perspective is organized as follows. First, various ferroelectric mechanisms found in 2D perovskite systems are introduced, involving proper ferroelectricity, octahedral distortion-induced ferroelectricity, and spin-induced ferroelectricity. Then, an overview of 2D perovskite multiferroics is presented, and novel magnetoelectric coupling mechanisms corresponding to different ferroelectricity are described. Finally, some suggestions and future directions for the research of 2D perovskite ferroelectrics and multiferroics are proposed.

\section{2D ferroelectricity}

\subsection{Proper ferroelectricity}

Efforts to achieve 2D ferroelectricity can be traced back to conventional perovskite ferroelectric thin films. \cite{Qiao2021} However, the problem of critical thickness, below which the ferroelectricity vanishes, has always been an obstacle. \cite{Junquera2003,Fong2004,Kim2005,Woo2007,Kolpak2005} The experimentally observed thickness limit is influenced by many competing intrinsic and extrinsic effects, including size effects, electrical boundary conditions, substrate strain conditions, interface and surface chemistry, and sample quality, etc. \cite{Qiao2021} Therefore, the reported thickness limit of perovskite ferroelectric films is continuously reduced and even absent. At present, the critical thickness of conventional perovskite ferroelectric films has generally been reduced to several unit cell thickness, such as 3 unit cells for BaTiO$_3$  \cite{Tenne2009,Shin2017} and PbTiO$_3$, \cite{Fong2004} and 1.5 unit cells for PbZr$_{0.2}$Ti$_{0.8}$O$_3$, \cite{Gao2017} which have entered the 2D region. Furthermore, theoretical work suggested that ferroelectricity can persist down to one unit cell with proper electrodes, \cite{Kolpak2005} and experimentally, OP ferroelectricity down to the monolayer limit has recently been demonstrated in tetragonal BiFeO$_3$ ultrathin films. \cite{Wang2018} In general, the performance of perovskite ferroelectric ultrathin films deteriorates with decreasing thickness, resulting in the saturation polarization and Curie temperature of 2D perovskite ferroelectrics lower than their bulk counterparts.\cite{Qiao2021} However, there are some exceptions, such as the 2D multiferroic BiFeO$_3$ film, which retains a tetragonal phase with an abnormally large $c/a$ ratio in its freestanding form,\cite{Ji2019} resulting in a saturation polarization ($140 \mu{\rm C/cm^2}$) much larger than that of its unstrained bulk phase and the known 2D vdW ferroelectrics. \cite{Qi2021,Wang2023b}

Recently, the fabrication of freestanding single-crystalline oxide perovskite films has provided great opportunities for realizing perovskite-based 2D ferroelectrics. \cite{Ricciardulli2021} Some typical freestanding oxide perovskite ultrathin films, including SrTiO$_3$ and BiFeO$_3$, have been prepared by using molecular beam epitaxy and selective etching techniques, that is, depositing a water-soluble sacrificial buffer layer before the growth of the target oxide perovskite. \cite{Lu2016,Hong2017,Ji2019,Ricciardulli2021} The freestanding perovskite films can be transferred to any desired substrate, creating opportunities for designing multifunctional perovskite-based electronic devices. In addition, since freestanding perovskite films are not affected by substrate strain and interface effects, this provides an unprecedented opportunity to study the intrinsic effects affecting 2D ferroelectricity. Experimentally, in freestanding BiFeO$_3$ films, giant tetragonality emerges when approaching to the 2D limit, and OP polarization is switchable even in films as thin as two unit cells, \cite{Ji2019} as shown in Fig. 1(a).

Lu \emph{et al.} systematically studied the ferroelectricity of 2D freestanding perovskite oxide films using first-principles calculations, and confirmed that the proper ferroelectricity in BaTiO$_3$ films can persist even when the thickness is reduced to two unit cells, \cite{Lu2017} as shown in Fig. 1(b). However, the OP ferroelectric polarization is suppressed by the depolarization field, resulting in the polarization along the in-plane (IP) direction. Interestingly, they found a novel ferroelectric phenomenon, IP ferroelectricity driven by the surface effect, which is independent of the cationic electronic configuration and enhances with decreasing film thickness, unlike the behavior of conventional ferroelectricity. Our work shows the absence of ferroelectricity in some freestanding perovskite monolayers with formula $A_2$$B$O$_4$ (\emph{A}= Ca, Sr; \emph{B} = Ti, Zr, Si, Ge, Sn). \cite{Zhou2022} Besides oxide perovskite ferroelectrics, the proper ferroelectricity driven by the lone pair electronic configuration ($ns^2$) of the \emph{B}-site ions exist in freestanding halide perovskite monolayers with smaller anions, such as the monolayers of CsGeF$_3$, CsSnF$_3$, and CsSnCl$_3$. \cite{Zhang2024} Therefore, the conventional ferroelectric mechanisms can generally exist in 2D perovskites, but the ferroelectricity may be different from that of the bulk phase due to structural reconstruction, and tends to form IP ferroelectricity in the absence of electrode to shield the depolarization field.

\subsection{Hybrid improper ferroelectricity}

\begin{figure*}
\centering
\includegraphics*[width=0.95\textwidth]{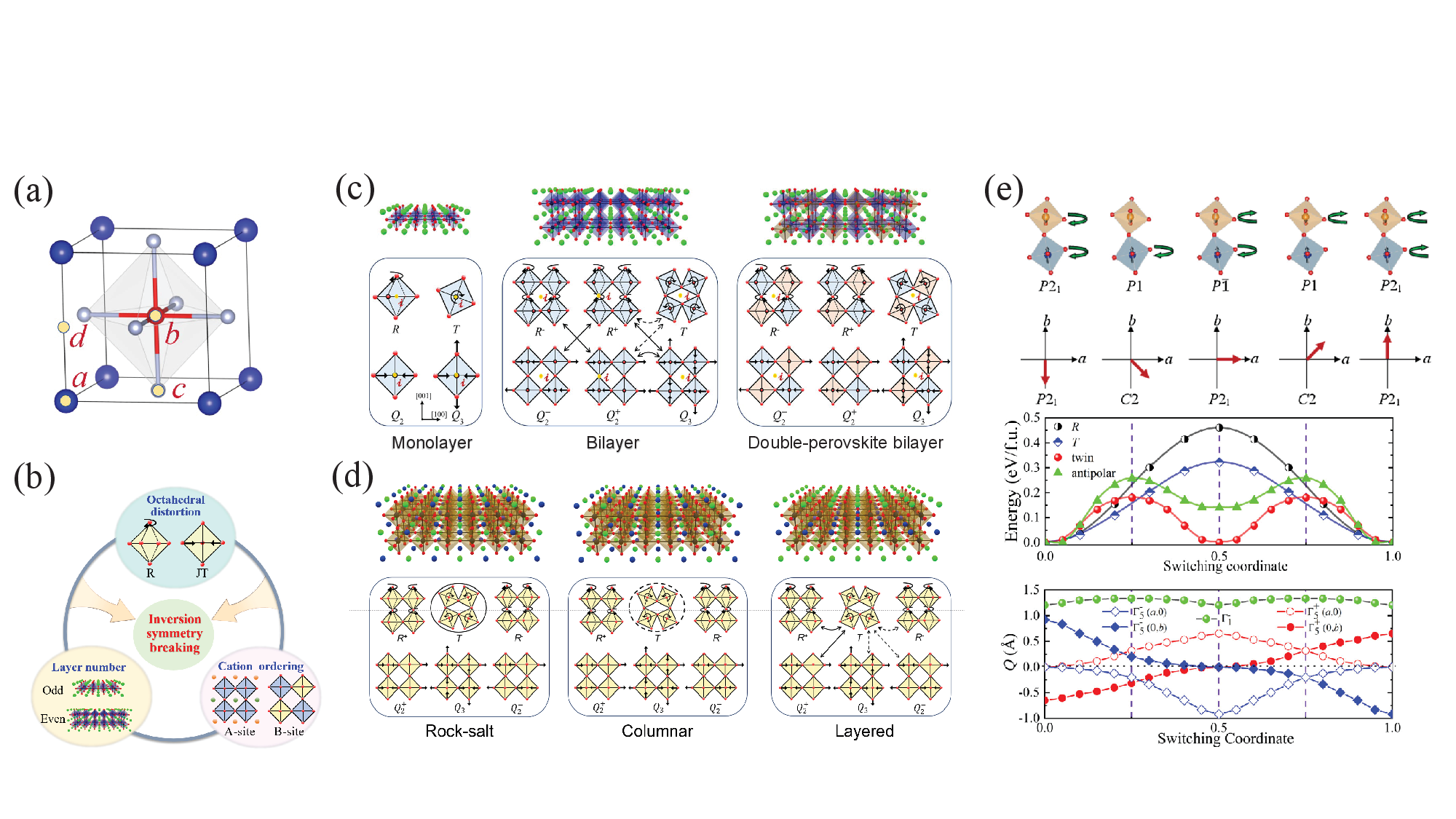}
\caption{\label{Fig2} (a) Crystal structure of the perovskite prototype phase and the four Wyckoff positions with inversion centers. (b) The common octahedral distortion in perovskites combined with layer number (2D perovskites) or cation order causes the inversion symmetry breaking. (c) Schematic octahedral rotation and JT distortion of perovskite monolayer, bilayer, and double-perovskite bilayer. The solid dot marked by the letter \emph{i} represents the inversion center existing in the distorted structure, and the solid and dashed double arrows indicate that the combination of the two modes leads to nonpolar and polar point groups, respectively. Reproduced with permission from Zhang \emph{et al.}, Phys. Rev. Lett., 129, 117603 (2022). Copyright 2022 American Physical Society. \cite{Zhang2022} (d) Ferroelectricity induced by octahedral distortion in \emph{A}-site ordered perovskites. The solid and dashed ellipses represent that the single tilt mode can cause ferroelectricity in any number of layers and in only even layers, respectively. The solid and dashed double arrows represent that the combinations of the two modes can cause ferroelectricity in any number of layers and in only odd layers, respectively. Reproduced with permission from Shen \emph{et al.}, Nano Lett., 23, 735 (2023). Copyright 2023 American Chemical Society. \cite{Shen2023} (e) Schematic and energy barrier of different ferroelectric switching paths in a perovskite bilayer with octahedral rotation-induced ferroelectricity.  Reproduced with permission from Zhang \emph{et al.}, Phys. Rev. Lett., 125, 017601 (2020). Copyright 2020 American Physical Society. \cite{Zhang2020}}
\end{figure*}
Hybrid improper ferroelectricity was first demonstrated in perovskite oxide superlattices, \cite{Bousquet2008} where the polarization results from its coupling effect with two octahedral rotation distortion modes (primary order parameters), known as the trilinear coupling. Subsequently, similar improper ferroelectricity was proposed in layered perovskite oxides of the Ruddlesden-Popper type, \cite{Benedek2011,Benedek2022} and then experimentally confirmed in $A_3$$B_2$O$_7$ (\emph{A} = Ca, Sr; \emph{B} = Ti, Mn, Zr, Sn). \cite{Oh2015,Huang2016,Gao2017a,Liu2018,Yoshida2018,Yoshida2018a} The octahedral rotation distortion arises from the structural geometry effect, which is related to the mismatch of cation radius at the \emph{A} and \emph{B} sites, and is therefore very common in perovskite system. \cite{Lufaso2004} Hybrid improper ferroelectricity, which relies only on a specific type of octahedral rotation, such as the $a^-a^-c^+$ type common in perovskites, and is independent of the cationic electronic configuration, is expected to be more widespread than conventional ferroelectricity. \cite{Benedek2011,Benedek2022}

However, common octahedral distortion in perovskites, including rotation and JT distortion, can break inversion centers at other sites (Wyckoff positions \emph{a}, \emph{c}, \emph{d}) but always retain the one at position \emph{b} [see Fig. 2(a)], thus ruling out the possibility of ferroelectricity induced by octahedral distortion in simple perovskite bulks. Therefore, this improper ferroelectricity requires the octahedral distortion combined with other degrees of freedom, such as the number of layers of 2D and layered perovskites, \cite{Benedek2011, Zhou2022} or the cationic order, \cite{Mulder2013,Rondinelli2012,Zhao2014,Shaikh2021,Lu2016a} to break the inversion symmetry, as shown in Fig. 2(b). For monolayer (or other odd-layer) perovskites with complete octahedral coordination, ferroelectricity induced by octahedral distortion does not occur because the inversion center at position \emph{b} is retained. In even-layer perovskites, the possible inversion centers shift to the middle \emph{A}\emph{X}$_2$ plane, involving the positions \emph{a}, \emph{c}, and \emph{d}. The in-phase octahedral distortion retains the inversion centers at positions \emph{a} and \emph{c}, while the out-of-phase distortion modes take position \emph{d} as the inversion center. Therefore, the combinations of two octahedral distortion modes with different inversion centers break the inversion symmetry, among which the combination of in-phase rotation or JT distortion with tilt modes establishes a polar point group, allowing the emergence of hybrid improper ferroelectricity, \cite{Zhang2022} as shown in Fig. 2(c). This ferroelectricity induced by octahedral rotation distortion is widespread in perovskite bilayer system \cite{Zhou2022}.

\begin{table}
\centering
\includegraphics*[width=0.48\textwidth]{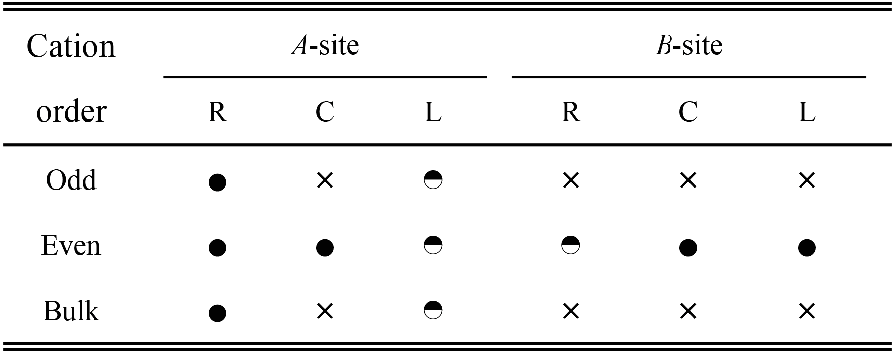}
\caption{\label{table 1} Ferroelectricity induced by octahedral distortion in perovskites with different cation orders and number of layers. The \emph{A}-site and \emph{B}-site cation orders considered involve rock-salt (R), columnar (C), and layered (L) orders. The number of perovskite layers is divided into odd, even and bulk. The fully filled and half-filled circle symbols represent that ferroelectricity can be induced by a single distortion mode or a combination of two modes, respectively, and the cross symbol indicates that ferroelectricity does not occur.}
\end{table}

Cationic order can enrich the ferroelectricity induced by octahedral distortion and facilitate the introduction of ferromagnetism (or ferrimagnetism) to achieve multiferroics. \cite{Shen2023,Zhang2020,Lu2016a} The most common cationic order is the \emph{B}-site order, usually of the rock-salt type, \cite{Vasala2015} while the \emph{A}-site order generally prefers the layered type. \emph{B}-site ordered configurations, including rock-salt, columnar, and layered orders, all retain the inversion center at position \emph{b}, resulting in the absence of octahedral distortion-induced ferroelectricity in odd-layer and bulk perovskites with \emph{B}-site order (see Table 1). Interestingly, a single octahedral distortion mode, such as tilt, can induce ferroelectricity in even-layer perovskites with columnar and layered orders. \cite{Shen2023} For the \emph{A}-site order, the rock-salt and layered types break the inversion center at position \emph{b}, so octahedral distortion-induced ferroelectricity can emerge in any number of layers (or even in the bulk) with these cationic orders. The difference is that in the former cationic order, ferroelectricity can be induced by a single tilt mode, whereas in the latter, it requires a combination of tilt and other distortion modes, as shown in Fig. 2(d).

Due to the trilinear coupling effect, switching either of the two octahedral distortion modes leads to a reversal of ferroelectric polarization, so the ferroelectric switching is accompanied by the transition of one of the octahedral distortion modes. Each octahedral distortion mode can be switched via one step or multistep, which undergo an intermediate phase without or with this mode, respectively. Our work demonstrates that the multistep switching path of the tilt mode generally has the lowest energy, which is accomplished by continuous IP rotation of the tilt axis and thus undergoes an orthogonal twin intermediate state, \cite{Zhang2020} as shown in Fig. 2(e). Such orthogonal twin domains have been demonstrated in hybrid improper ferroelectric Ca$_3$Ti$_2$O$_7$, \cite{Oh2015,Huang2016} which are responsible for its low switching barrier. In general, a larger perovskite tolerance tends to form non-ferroelectric phase due to the disappearance of octahedral rotation or tilt mode, while a smaller tolerance factor will result in a high energy barrier that is not conductive to ferroelectric switching. Therefore, the perovskite tolerance factor plays a key role in this improper ferroelectricity.

\subsection{Spin-induced ferroelectricity}

\begin{figure}
\centering
\includegraphics*[width=0.45\textwidth]{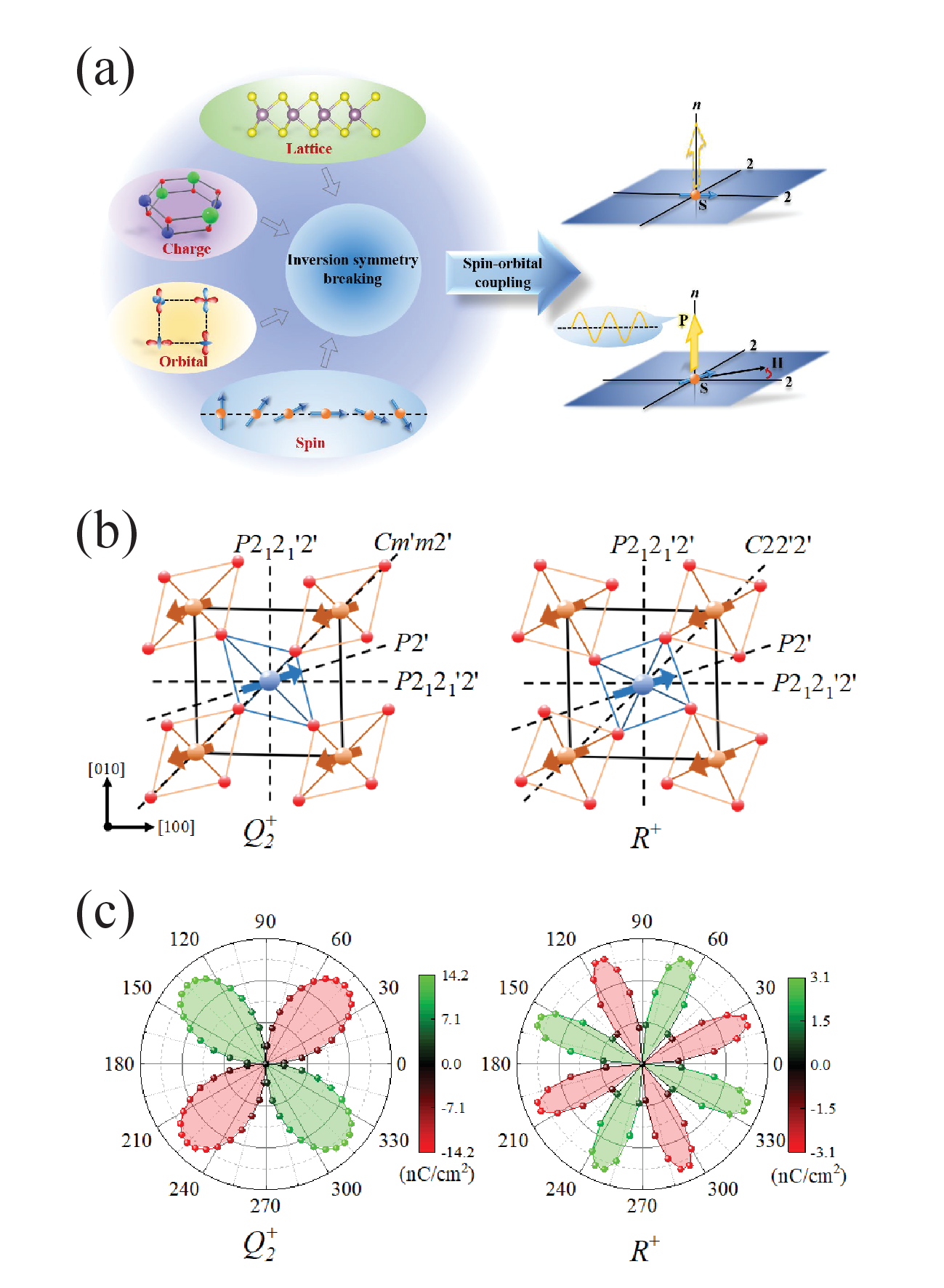}
\caption{\label{Fig3} (a) Schematic of spin-direction-dependent ferroelectricity induced by the single spin effect. Dependence of (b) magnetic symmetry and (c) OP ferroelectric polarization on the IP spin direction in the distorted structures caused by the in-phase JT (left) and rotation (right) modes individually in the \emph{B}-site rock-salt ordered perovskite bilayers. The two types of arrows in different colors represent the spins of the two magnetic ions. Reproduced with permission from Zhang \emph{et al.}, Phys. Rev. Lett., 129, 117603 (2022). Copyright 2022 American Physical Society \cite{Zhang2022}.}
\end{figure}
Spin-induced ferroelectricity in perovskites was first observed in TbMnO$_3$, \cite{Kimura2003} a typical type-II multiferroic material, whose ferroelectricity originates from the spiral spin order and can be explained by the spin-current mechanism. \cite{Katsura2005}. Another common mechanism for spin-induced ferroelectricity, the exchange-striction mechanism, usually occurs in complex collinear antiferromagnetic structures, such as the \emph{E}-type spin order appearing in perovskites \emph{R}MnO$_3$ (\emph{R}= Ho-Lu). \cite{Sergienko2006} The third mechanism, \emph{p-d} hybridization involving spin-orbit coupling (SOC), \cite{Murakawa2010} has never occurred in perovskite bulk due to the inversion symmetry in octahedral coordination.

Spin-induced ferroelectricity usually does not occur in perovskite materials with simple magnetic structures, since it requires complex spin order to break spatial inversion symmetry. In fact, however, spin-induced ferroelectricity can occur in simple antiferromagnetic or even ferromagnetic orders if inversion symmetry is broken by other degrees of freedom, such as lattice, charge, and orbital orders. Specifically, a single spin, which acts as an axis vector under SOC, remains unchanged under inversion symmetry, but can break the nonpolar symmetry to polar symmetry, resulting in spin-direction-dependent ferroelectricity, as shown in Fig. 3(a). Therefore, inversion symmetry breaking and SOC are prerequisites for the emergence of spin-induced ferroelectricity in simple collinear magnetic structures. \cite{Zhang2022}

Our work demonstrates that this spin-induced ferroelectricity can exist in 2D perovskites with simple magnetic structures, including ferromagnetic and ferrimagnetic orders. As mentioned above, inversion symmetry in 2D perovskites can be broken by octahedral distortion, which can be accomplished by a single distortion mode when combined with cationic order. \cite{Zhang2022} In \emph{B}-site ordered perovskite bilayer, both in-phase rotation and JT distortion individually establish nonpolar structural phases [see Fig. 2(c)]. When the spin orientation changes within the \emph{ab} plane, their magnetic symmetry transitions from nonpolar to polar, allowing the emergence of OP ferroelectricity, as shown in Fig. 3(b). The dependence of ferroelectric polarization on the spin direction was confirmed by first-principles calculations, which show different modulation periods for the two structural phases [see Fig. 3(c)]. This ferroelectricity can be explained by modifying the \emph{p-d} hybridization mechanism, that is, taking into account the mutual influence of  \emph{p-d} hybridization of different coordination bonds and even higher-order perturbative effects of SOC. The actual spin-induced polarization and its compatibility with ferromagnetism and ferrimagnetism have been theoretically demonstrated in some 2D perovskite oxides. \cite{Zhang2022}

\section{2D multiferroics and magnetoelectric coupling}

Most known multiferroics materials are antiferromagnetic and lack intrinsic magnetoelectric coupling, so achieving the coexistence and coupling of ferroelectricity and ferromagnetism above room temperature remains the primary challenge in the research on multiferroics. \cite{Eerenstein2006,Spaldin2019} These novel ferroelectric mechanisms discovered in 2D perovskite systems may provide an effective approach to address this challenge.

Reducing known perovskite multiferroic materials to the 2D limit may be a direct route to 2D multiferroics. For example, the ferroelectricity of the typical multiferroic BiFeO$_3$ has been shown to be retained to the 2D limit, although its magnetic property in 2D has not been revealed. \cite{Ji2019,Wang2018} Unlike vdW materials, perovskite materials may undergo change in chemical composition or even structural reconstruction when reduced to 2D, resulting in changes in ferroelectric and magnetic properties. On the other hand, the reduction in symmetry relative to the perovskite bulks may lead to the emergence of some novel multiferroic phenomena and magnetoelectric coupling mechanisms in 2D perovskites.

\subsection{Controlling magnetization by electric field}

\begin{figure}
\centering
\includegraphics*[width=0.45\textwidth]{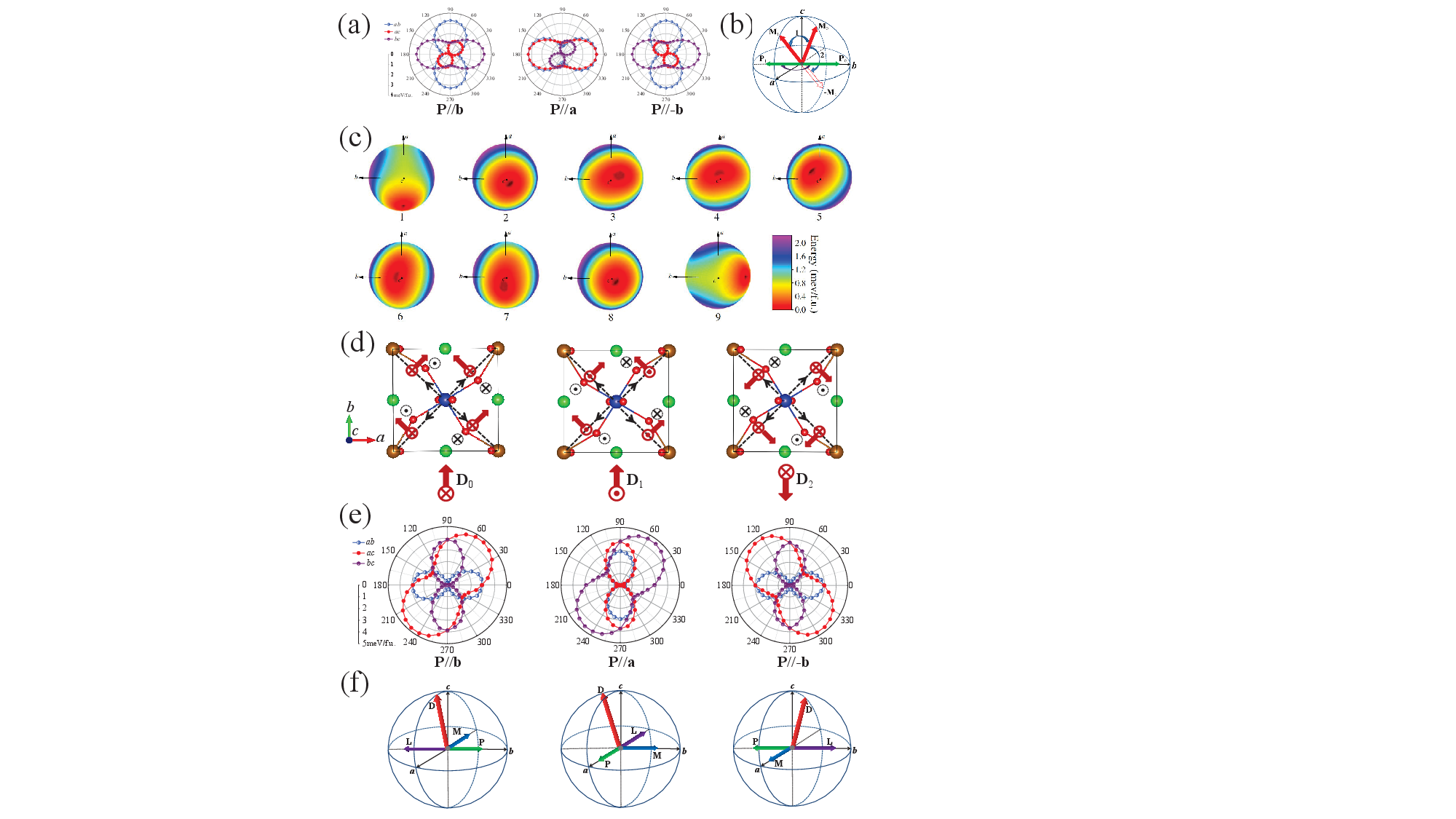}
\caption{\label{Fig4} (a) Magnetic anisotropy energy of the initial, orthogonal twin, and final states in ferroelectric switching of 2D perovskite multiferroic Ca$_3$FeOsO$_7$. (b) Schematic of the change in the magnetization direction with the reversal of polarization. (c) Evolution of magnetic anisotropy energy surface from the initial state to the orthogonal twin state. (d) The orientation of the DM vectors of IP bonds. The middle and right panels represent their transformations caused by the reversal of rotation and tilt distortion, respectively. (e) Magnetic anisotropy energy including DM interaction and (f) schematic of the orientations of the total DM vector (\textbf{D}), antiferromagnetic vector (\textbf{L}), and net magnetization (\textbf{M}) for the initial, intermediate, and final states. Reproduced with permission from Zhang \emph{et al.}, Phys. Rev. Lett., 125, 017601 (2020). Copyright 2020 American Physical Society. \cite{Zhang2020}}
\end{figure}

Multiferroics can be extensively achieved in 2D perovskite system by combining ferroelectricity induced by octahedral rotation distortion and ferromagnetism (or ferrimagnetism) induced by \emph{B}-site cationic order. \cite{Zhang2020,Shen2021} Importantly, the ferroelectric polarization and macroscopic magnetization are coupled via the octahedral rotation distortion. When ferroelectric polarization is reversed along the lowest energy switching path, that is, reversing tilt mode via a multi-step switching, the easy-magnetization axis shifts to a direction symmetric about the \emph{c} axis with respect to its initial state, as shown in Fig. 4(a). This magnetoelectric coupling phenomenon in which ferroelectric switching causes a change in the easy-magnetization axis is similar to that of BiFeO$_3$. However, the difference is that in BiFeO$_3$, a 180$^\circ$ switch in ferroelectric polarization does not change the direction of the easy-magnetization axis. \cite{Zhao2006} The evolution of the easy-magnetization axis during ferroelectric switching indicates that the IP magnetization component is reversed while the OP component remains unchanged [see Figs. 4(b) and 4(c)]. The transition of magnetization direction caused by ferroelectric switching is protected by symmetry, that is, the symmetry correlation between the structures of the initial and final polarization states, and is therefore robust. \cite{Shen2021} Similar magnetoelectric coupling effects also exist in perovskite monolayer multiferroics with \emph{A}-site cationic order. \cite{Shen2023}

Another magnetoelectric coupling mechanism, which relies on Dzyaloshinskii-Moriya (DM) interaction, also exists in 2D perovskite systems. \cite{Zhang2020} DM interaction can induce spin canting in antiferromagnetic structures, resulting in weak ferromagnetism. In some double-perovskite oxides, DM interaction can induce a considerable net magnetic moment due to the strong SOC of the 5\emph{d} ions. Furthermore, the direction of the DM vectors is determined by the octahedral rotation distortion. Switching the rotation and tilt modes will result in the reversal of the OP and IP components of the DM vectors of the IP bonds, respectively, as shown in Fig. 4(d). Therefore, in 2D perovskite multiferroic materials where ferroelectricity and ferromagnetism are derived from octahedral rotation distortion and DM interaction, respectively, ferroelectric polarization and net magnetization are coupled by means of DM interaction. During the ferroelectric switching, the continuous rotation of the octahedral tilt axis in the \emph{ab} plane causes the total DM vector to rotate around the \emph{c} axis, which results in the simultaneous reversal of the antiferromagnetic vector (spin direction) and the ferromagnetic component when approaching the final polar state [see Figs. 4(e) and 4(f)]]. This is different from the previously proposed magnetoelectric coupling mechanism related to DM interaction, in which the DM vector is reversed while the antiferromagnetic vector remains unchanged in ferroelectric switching. \cite{Fennie2008} These two novel magnetoelectric coupling mechanisms in 2D perovskite multiferroics enable the reversal of the IP magnetization by electric field, as shown in Figs. 5(a) and 5(b).

\subsection{Controlling polarization by magnetic field}

\begin{figure}
\centering
\includegraphics*[width=0.45\textwidth]{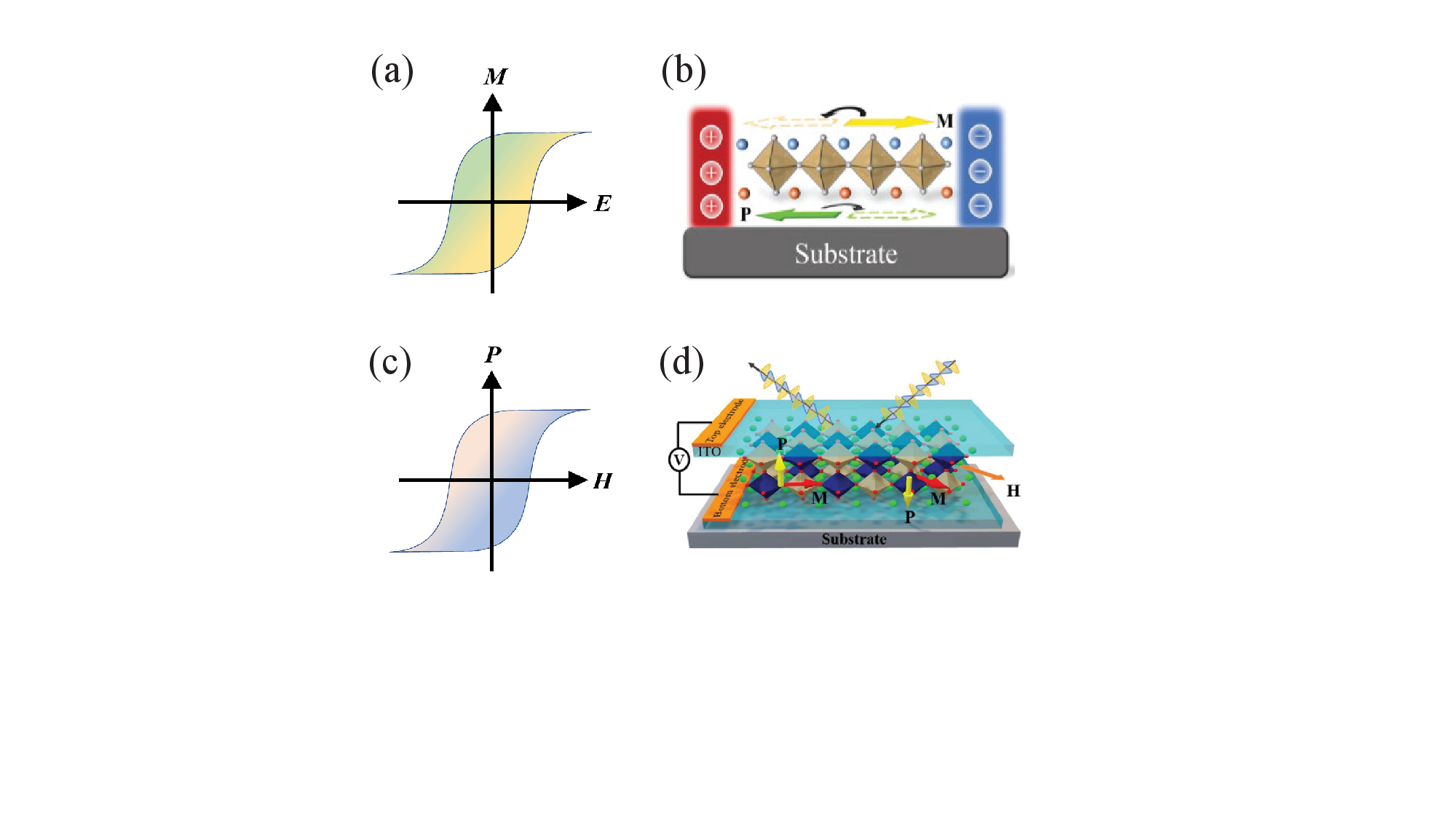}
\caption{\label{Fig5} (a) Schematic of controlling magnetization using electric field and (b) corresponding magnetoelectric coupling in 2D perovskite multiferroic with octahedral distortion-induced ferroelectricity. Reproduced with permission from Shen \emph{et al.}, Nano Lett., 23, 735 (2023). Copyright 2023 American Chemical Society. \cite{Shen2023} (c) Schematic of controlling polarization using magnetic field and (d) corresponding magnetoelectric coupling in 2D perovskite multiferroic with spin-induced ferroelectricity. Reproduced with permission from Zhang \emph{et al.}, Phys. Rev. Lett., 129, 117603 (2022). Copyright 2022 American Physical Society. \cite{Zhang2022}}
\end{figure}

Controlling ferroelectric polarization by magnetic field may be achieved in type-II multiferroics with spin-induced ferroelectricity. However, such multiferroic materials are commonly found in antiferromagnets, so their spin-induced polarization is difficult to be effectively controlled by a magnetic field. \cite{Cheong2007,Tokura2010,Tokura2014} By means of the mechanism of spin-induced ferroelectricity mentioned above, not only can spin-induced ferroelectricity and ferromagnetism coexist in 2D perovskites, but also the control of polarization by magnetic field can be achieved, as shown in Figs 5(c) and 5(d). The dependence of polarization on spin direction indicates that changing the spin direction not only modulate the magnitude of the OP polarization, but also reverse its direction [see Fig. 3(c)]. Therefore, when an external magnetic field is applied to overcome the magnetic anisotropy energy, the spin direction and the induced ferroelectric polarization will change simultaneously with the direction of the magnetic field. \cite{Zhang2022}

Note that octahedral distortion-induced ferroelectricity and spin-induced ferroelectricity can occur in the same 2D perovskite materials, so the related magnetoelectric cross-control effects may exist simultaneously.\cite{Zhang2022} In addition to perovskites, similar spin-induced ferroelectricity also occurs in 2D vdW ferromagnets that lack inversion symmetry, where the induced polarization can reach the maximum known for type-II multiferroics. \cite{Wang2023,Zhou2024} Furthermore, the cross-control effects of external electric and magnetic fields on different ferroic order parameters have been theoretically demonstrated in these 2D multiferroic materials.

\section{Future aspects}

\subsection{OP ferroelectricity}

For 2D ferroelectrics, OP ferroelectricity is more desirable for high-density nanodevices, but also more challenging. At present, OP ferroelectricity has been demonstrated in some 2D vdW ferroelectrics, \cite{Qiao2021} but it rarely occurs in pristine 2D perovskite films. Although the OP ferroelectricity in BiFeO$_3$ can be retained to the 2D limit, the interface effect with the SrRuO$_3$ electrode plays a key role in stabilizing the OP ferroelectricity at atomic thickness. \cite{Wang2018} Whether other conventional perovskite ferroelectrics exhibit switchable OP ferroelectricity in the 2D limit with the assistance of external effects such as proper electrodes remains to be further studied. In addition, theoretical studies have shown that most conventional perovskite ferroelectrics exhibit IP ferroelectricity in their freestanding 2D films, so whether OP ferroelectricity can exist in pristine 2D perovskite films remains an open question. Note that 2D perovskite ferroelectrics may exhibit a critical thickness beyond which a transition between IP and OP ferroelectricity occurs.

As for improper ferroelectricity, all octahedral distortion-induced ferroelectricity found in 2D perovskites exhibits IP ferroelectricity. \cite{Zhou2022,Zhou2021,Shen2023,Zhang2020,Shen2021} Compared with the proper ferroelectricity, ferroelectricity induced by octahedral distortion may be more stable against the influence of depolarization filed because it originates from the structural geometry effect. Theoretically, OP ferroelectricity can be caused in 2D perovskite systems through specific combinations of octahedral distortion modes such as in-phase plus out-of-phase rotation, or special octahedral distortion combined with cationic order, such as $Q_3$ type of JT distortion in \emph{B}-site columnar ordered perovskite bilayers. \cite{Zhang2022} However, these specific combinations required for OP ferroelectricity have not yet been found in actual 2D perovskite materials. Spin-induced OP ferroelectricity can exist in perovskite bilayers lacking inversion symmetry, but the induced polarization is much smaller than those of conventional perovskite ferroelectrics. Therefore, the polarization of such ferroelectrics needs to be significantly improved before their application in nanoelectronic devices.

\subsection{Ferroelectric switching}

\begin{figure*}
\centering
\includegraphics*[width=0.9\textwidth]{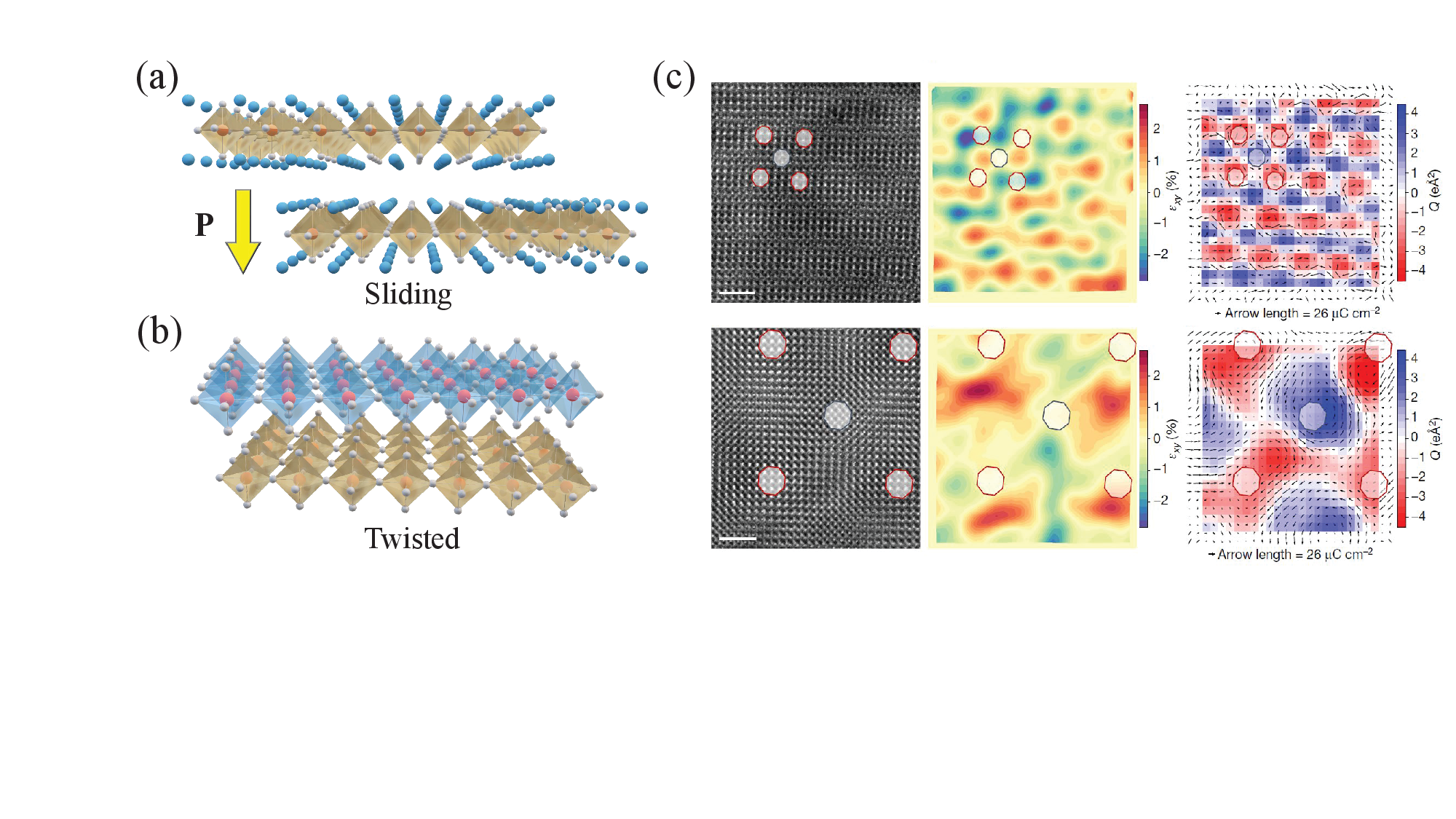}
\caption{\label{Fig6} (a) Schematic of ferroelectric polarization induced by interlayer sliding in stacked perovskite bilayers. (b) Stacked perovskite bilayers with twist angles. (c) Strain and polarization modulations at twisted freestanding BaTiO$_3$ bilayers, show a network of clockwise and anticlockwise votices. Reproduced with permission from S\'{a}nchez-Santolino \emph{et al.}, Nature, 626, 529 (2024). Copyright 2024 Springer Nature. \cite{SanchezSantolino2024}}
\end{figure*}

Ferroelectric switching of 2D BiFeO$_3$ films has been demonstrated by high-resolution piezoresponse force microscopy in different experimental studies, showing that the switchable OP ferroelectricity can be retained in monolayer and bilayers in epitaxial strained and freestanding films, respectively. \cite{Ji2019,Wang2018} Therefore, the polarization switching behavior of some other conventional ferroelectrics when approaching the 2D limit and its important influencing factors deserve further study.

For ferroelectricity induced by octahedral rotation distortion, its ferroelectric switching has been observed in layered perovskite oxides, \cite{Oh2015,Huang2016,Gao2017a,Liu2018} but has not yet been experimentally demonstrated in 2D perovskites. There are multiple possible ferroelectric switching paths for this ferroelectricity, \emph{i.e.}, reversing the rotation or tilt mode? Via one step or multiple steps? Importantly, the magnetoelectric coupling effect is determined by the ferroelectric switching path, specifically, only the switching of the tilt mode results in a change in the magnetization direction. \cite{Shen2021} At present, the theoretical approach to determine ferroelectric switching path of such ferroelectrics is usually to compare the energy barriers of different ferroelectric switching paths through first-principles calculations. \cite{Zhou2022,Zhou2021,Shen2023,Zhang2020,Shen2021,Nowadnick2016} However, this method only simulates the switching of homogeneous polarization in a single domain state, while the actual ferroelectric switching involves the dynamic evolution of domain structure and inhomogeneous polarization. Therefore, more accurate theoretical methods such as phase field simulation are needed to study the ferroelectric switching process. Experimentally, the ferroelectric switching path can be determined by observing the structures of the initial, final and intermediate polarization states using high-resolution scanning transmission electron microscopy.

The predicted spin-induced ferroelectricity in 2D perovskites and its switchability under electric field also need experimental confirmation. Due to the dependence of polarization on spin direction, the switching of polarization will inevitably lead to a change in magnetization direction. Therefore, it is necessary to study the evolution of polarization and spin direction under external electric field to reveal the control effect of electric field on magnetism in such multiferroic materials.

\subsection{Novel ferroelectric and multiferroic phenomena}
Whether there are other novel ferroelectric and magnetoelectric coupling mechanisms in 2D perovskite systems deserves further experimental and theoretical studies. Draw on the interesting ferroelectric phenomena found in 2D vdW materials is an effective way to discover novel ferroelectricity in 2D perovskites. Ferroelectricity induced by interlay stacking and sliding is a typical and common ferroelectric phenomenon in vdW materials, which has been experimentally confirmed. \cite{ViznerStern2021,Yasuda2021,Wang2022,Rogee2022,Cao2018} OP polarization can be induced by stacking non-ferroelectric monolayers into multilayers, and it can be switched by interlay sliding. The ability to prepare and transfer 2D freestanding perovskite films enables stacking of perovskite monolayer to obtain homogeneous multilayers and heterostructures [see Fig. 6(a)], as has been done for vdW materials. However, unlike vdW materials, the unsaturated bonds on the terminal surfaces of 2D perovskites leads to stronger interlayer interactions, so the resulting sliding ferroelectricity will have a larger ferroelectric polarization and a higher switching energy barrier than vdW materials. In theory, the general rules of sliding ferroelectricity in 2D perovskites, such as the requirement of ferroelectricity on the symmetry of perovskite monolayer and the stacking configurations, can be established by symmetry analysis, and then the actual perovskite materials that meet the requirement for ferroelectricity can be screened out using first-principles calculation.

In recent years, since the discovery of unconventional superconductivity in magic-angle twisted graphene bilayers, twisted vdW multilayers have become an important platform for exploring exotic electronic phases and emergent phenomena, resulting in the development of a new branch of twistronics. \cite{Cao2018,Sharpe2019,Andrei2020,Ciarrocchi2022,Carr2017} Compared with vdW materials, transition-metal perovskites exhibit richer correlated electronic behaviors, so more exotic correlated interfacial phases are yet to be explored in twisted perovskite multilayers [see Fig. 6(b)]. Whether novel ferroelectric and multiferroic phenomena emerge in twisted perovskite multilayers also deserves further study. \cite{Zheng2020,Han2023} A recent experimental study demonstrates the ability to create non-trivial ferroelectric textures based on twisted freestanding perovskite ferroelectric layers. \cite{SanchezSantolino2024} As shown in Fig. 6(c), a network composed of clockwise and anticlockwise vortices was observed in twisted BaTiO$_3$ bilayers, which arises from the flexoelectric coupling of polarization to strain gradients, that is, the inhomogeneous strain distribution imposed by the interface causes the vortex-like modulation of polarization through the flexoelectric effect. This work demonstrates the great potential of stacked perovskite layers with twist angles in designing novel ferroelectric textures and exploring unknown physical effects.

\begin{acknowledgments}
This work was supported by the National Natural Science Foundation of China (Grants No. 12374097 and No. 11974418), and the Postgraduate Research \& Practice Innovation Program of Jiangsu Province (grant number KYCX24\_2693).
\end{acknowledgments}

\section*{Author declarations}
\subsection*{Conflict of Interest}
\quad The authors have no conflicts to disclose.

\subsection*{Author Contributions}
\noindent\textbf{Junting Zhang:} Writing original draft (lead). \textbf{Yu Xie:} Writing review \& editing (equal). \textbf{Ke Ji:} Writing review \& editing (supporting). \textbf{Xiaofan Shen:} Writing review \& editing (equal)

\section*{Data availability}

Data sharing is not applicable to this article as no new data were
created or analyzed in this study.

%

\end{document}